\RequirePackage{lineno}
\documentclass[aps,twocolumn,showpacs,preprintnumbers,amsmath,amssymb]{revtex4}
\usepackage{graphicx}% Include figure files 
\usepackage{dcolumn}% Align table columns on decimal point 
\usepackage{epstopdf}
\usepackage{color}
\usepackage{amssymb}
\usepackage{bm}% bold math 
\usepackage{subfigure}
\definecolor{dgreen}{cmyk}{1.,0.,1.,0.2}        % dark green 
\definecolor{orange}{cmyk}{0.,0.353,1.,0.}    % orange 

\def\bea {\begin{eqnarray}}
\def\eea {\end{eqnarray}}

\def\be {\begin{equation}}
\def\ee {\end{equation}}

\usepackage{hyperref}% to enable the internal references 
\RequirePackage{lineno}

\def\bea {\begin{eqnarray}}
\def\eea {\end{eqnarray}}

\def\be {\begin{equation}}
\def\ee {\end{equation}}

\newcommand{\CV}{$C$}
\newcommand{\cv}{$c_{\rm v}$}
\newcommand{\cvT}{$C/(VT^3)$}
\newcommand{\cvN}{$C/N$}

\newcommand{\Teff}{$T_{\rm eff}$}
\newcommand{\Tkin}{$T_{\rm kin}$}
\newcommand{\Tch}{$T_{\rm ch}$}
\newcommand{\meanpt}{$\langle p_{\rm{T}} \rangle$}

\newcommand{\pT}{$p_{\rm{T}}$}
\newcommand{\pt}{$p_{\rm{T}}$}

\newcommand{\sNN}{$\sqrtsign {s_{\rm NN}}$}

\begin{document}
%\linenumbers

\title{Specific Heat of Matter Formed in Relativistic Nuclear Collisions}

\author{Sumit Basu}\affiliation{Variable Energy Cyclotron Centre, Kolkata 700064, India}
\author{Sandeep Chatterjee}\affiliation{Variable Energy Cyclotron Centre, Kolkata 700064, India}
\author{Rupa Chatterjee}\affiliation{Variable Energy Cyclotron Centre, Kolkata 700064, India}
\author{Basanta K. Nandi}\affiliation{Indian Institute of Technology Bombay, Mumbai 400076, India}
\author{Tapan K. Nayak}\affiliation{Variable Energy Cyclotron Centre, Kolkata 700064, India}

\bigskip

\date{ \today}

\begin{abstract}

We report the excitation energy dependence of specific heat (\cv) 
of hadronic matter at freeze-out in Au+Au and Cu+Cu collisions 
at the Relativistic Heavy Ion Collider energies by 
analyzing the published data on event-by-event mean transverse 
momentum (\meanpt) distributions. The \meanpt~distributions in 
finite \pt~ranges are converted to distributions of effective 
temperatures, and dynamical fluctuations in temperature are extracted by subtracting 
widths of the corresponding mixed event distributions. The heat capacity 
per particle at the kinetic freeze-out surface is presented as a 
function of collision energy, which 
shows a sharp rise in \cv~below \sNN~=~62.4~GeV.
We employ the Hadron Resonance Gas 
(HRG) model to estimate \cv~at the chemical and kinetic freeze-out 
surfaces.  The experimental results are compared to the HRG and other 
theoretical model calculations. HRG results show good agreement with 
data. Model predictions for \cv~at the Large Hadron Collider energy
are presented.

\end{abstract}
\pacs{25.75.-q,25.75.Gz,25.75.Nq,25.75.Dw,12.38.Mh}
\maketitle

\section{Introduction}

The major 
goal of colliding heavy-ions at the 
Relativistic Heavy Ion Collider (RHIC) at Brookhaven National Laboratory and the 
Large Hadron Collider (LHC) at CERN 
is to study matter at
extreme conditions of temperature and energy densities, where  
quarks and gluons, rather than 
% nucleons and mesons
mesons and baryons, define the relevant
degrees of freedom~\cite{Aoki}. This new phase of matter, the Quark-Gluon Plasma
(QGP), is governed by the principles of Quantum Chromodynamics (QCD) and 
is the result of a phase transition from the normal nuclear
matter~\cite{intro1,intro2}. Experiments at RHIC and LHC are
on the quest to unearth the nature of the QCD phase transition 
and to get a glimpse of how matter behaves at extreme conditions. 
The beam energy scan (BES) program at RHIC has been initiated to
explore the onset of phase transition by scanning the collision
energy over a larger range and to locate the critical
point in the QCD phase diagram.

The thermodynamic state of the QCD matter can be 
specified by the temperature $T$ and the chemical potentials $\mu_B$, $\mu_S$ 
and $\mu_Q$  corresponding to the conserved charges of QCD, namely baryon 
number ($B$), strangeness ($S$), and electric charge ($Q$), respectively.
Phase transitions are associated with the transformation of 
thermodynamic quantities such as pressure, entropy and
energy density, as well as a set of response functions, like,
specific heat, compressibility and susceptibility with change in 
$T$, $\mu_B$, $\mu_Q$ and $\mu_S$. 
In this article, we discuss the specific heat (\cv) of the system produced 
in heavy-ion collisions at relativistic energies and its behaviour as 
a function of collision energy. 

Specific heat is a thermodynamic quantity
characterizing the equation of state of the system. For a system
undergoing phase transition, \cv~is expected to diverge at the critical point. 
Temperature fluctuation of the system provides an estimation of
\cv. Near the critical point, the specific heat is normally expressed in
terms of a power law, $c_{\rm v} \propto |T-T_c|^-\alpha$, where $T_c$ is
the critical temperature and $\alpha$ is critical exponent.
Thus the variation of thermal fluctuations with temperature
can be effectively used to probe the critical point.

%Because of the strong temperature dependence of \cv, 
%it can be used to test the properties of matter at thermal freeze-out. 
%Production of large number of particles in each
%collision at the RHIC and LHC energies makes it possible to study
%several quantities on an event-by-event 
%basis~\cite{rajagopal,steph1,steph2,heiselberg,sto,shuryak}. 
%This scenario has made it possible to assign a temperature to each
%event in heavy-ion collisions~\cite{sto,shuryak}.

Hadron Resonance Gas (HRG) model analysis of the particle yields 
indicate the formation of a thermal source for the produced particles in 
heavy-ion collisions~\cite{sto,shuryak}. 
The production of large number of particles in each collision at the RHIC 
and LHC energies makes it even possible to study several quantities on an 
event-by-event basis~\cite{rajagopal,steph1,steph2,heiselberg,sto,shuryak} and 
hence measure their event to event fluctuations. Thus, with the measurement of 
$T$ on an event-by-event basis, it is possible to extract the \cv~of the 
hot and dense strongly interacting matter produced in heavy-ion collisions. Assuming complete 
thermal equilibrium up to the surface of last scattering which is the 
kinetic freeze-out surface, \cv~is then expected to reveal the thermodynamic 
state of the matter at the moment of kinetic freeze-out.

The specific heat and its variation as a function of temperature have
been studied extensively in different theoretical
calculations. Statistical and HRG
models have been used to 
obtain \cv~as a function of
temperature in hadron gas and in quark-gluon
matter~\cite{sandip,ben-hao,X-M-Li,aram}. 
In Ref.~\cite{ben-hao}, a parton and hadron cascade model has been
used to investigate \cv~as a function of beam energy 
for the initial partonic stage using quark-gluon
matter and the final stage using hadronic matter.
Lattice QCD calculations~\cite{gavai,swagato,hatta} 
provide estimations of \cv~for a wide range of temperatures.
In Ref.~\cite{gavai}, 
continuum limits of \cv~have been calculated 
in quenched QCD at temperatures of  2$T_c$
and 3$T_c$, where $T_c$ is the transition temperature. It is found
that \cv~differs significantly from that of the ideal gas. Recent
lattice calculations using (2+1)-flavor QCD with almost physical quark
masses give the results of \cv~for a temperature range of 130 to 400
MeV~\cite{swagato}. 
% Results of the lattice calculations agree well with the HRG results
%at low temperatures. 
The low temperature (hadron phase) 
results agree well with HRG.

The specific heat has its 
origin in the event-by-event temperature fluctuations, which manifests
through the fluctuations in the transverse momenta
 (\pt)~\cite{sto,shuryak,aram,X-M-Li,korus,ben-hao,voloshin,Broniowski,Bozek,Bleicher}.
Event-by-event fluctuations of \meanpt~have been reported by
experiments at the CERN
Super Proton Synchrotron (SPS)~\cite{NA49-1,NA49-2,NA49-3,CERES} and
beam energy scan at RHIC~\cite{phenix,star1,star2,star_system}.
The values of \cv~extracted from the experimental 
results have large errors~\cite{korus,NA49-2,NA49-3,mike1}.
The \pT~fluctuation data from Ref.~\cite{NA49-1} yielded the value of
\cv~to be $60\pm 100$ at $T=180$~MeV for SPS energies. 
% 
% The goal of the present work is to derive specific heat from event-by-event
% temperature fluctuations at the freeze-out, that is
% thermodynamic fluctuations which occur after the elapse of sufficient
% time after the collision. 
% In this method, temperature fluctuations are deduced from 
% the \meanpt~distributions of charged particles. 
The statistical fluctuations
arising from the finite multiplicity distributions of charged
particles may significantly affect the extracted thermodynamic
fluctuations~\cite{voloshin}. In the present work, this is taken care 
of by subtracting the widths of the results of mixed events from the 
real data. Since radial flow affects the
estimation of temperature, its effect has also been considered. Finally,
the values of \cv~have been calculated as a function of beam energy
from published experimental data and compared to lattice and HRG
calculations. Further predictions have been made for the LHC energies. 

The paper is organized as follows. In Section-II, we present the
methodology for extraction of \cv~from event-by-event 
average transverse momentum distributions and from theoretical calculations.
The HRG model calculations of specific heat are presented in Section~III.
Event-by-event distributions of mean transverse momenta have been 
reported by several experiments at CERN-SPS and RHIC. The \meanpt~distributions and
corresponding values of effective temperatures 
are presented in Section~IV.
A compilation of the results of specific heat from
the existing experimental data
is presented in Section V. Estimates for the specific heat at LHC
using the AMPT model~\cite{ampt} have been included in this section. 
A discussion on the results of \cv is given in Section~VI.
The paper is summarised with an outlook in Section~VII.

\section{Specific heat: Methodology}

The heat capacity $C$ of a system is defined as~\cite{landau}:
% \begin{eqnarray}
% C_{\rm v} =  T \biggl(  \frac{\partial S}{\partial T} \biggr)_V = \biggl(  \frac{\partial E}{\partial T} \biggr)_{V,N} 
% \label{eqn1}
% \end{eqnarray}
% where $T$, $V$, $N$, $S$ and $E$ are temperature, volume, number of
% particles, entropy and energy of the system, respectively.
\begin{eqnarray}
%C_{\rm v} = \biggl(  \frac{\partial E}{\partial T} \biggr)_{V} 
C = \biggl(  \frac{\partial E}{\partial T} \biggr)_{V} 
\label{eqn1}
\end{eqnarray}
% where $T$, $V$, $N$, $S$ and $E$ are temperature, volume, number of
% particles, entropy and energy of the system, respectively. 
%Equivalently, \CV~can be expressed as the ratio of the event-by-event
%fluctuations of the energy ($E$) of a part of a finite system in
%thermal equilibrium to the energy:
where $T$, $V$ and $E$ are temperature, volume and energy of the system, respectively.
Equivalently, \CV~of a system in thermal equilibrium to a bath at $T$ can be computed 
from the event-by-event fluctuations of $E$:
\begin{eqnarray}
C =  
\frac{(\langle E^2 \rangle -  \langle E \rangle ^2 )}  
    { \langle T \rangle ^2}.
\label{eqn2}
\end{eqnarray}
%For a system in equilibrium, the transverse momentum spectra of
%emitted particle can be
%expressed in terms of event-by-event temperature fluctuation:
For a system in equilibrium, the event-by-event 
temperature fluctuation is controlled by the heat capacity:
\begin{eqnarray}
%P(T) \sim \exp [-\frac{C_{\rm v}}{2}  \frac{(\Delta T)^2}{\langle T
%\rangle^2}],
P(T) \sim \exp [-\frac{C}{2}  \frac{(\Delta T)^2}{\langle T \rangle^2}],
\label{eqn3}
\end{eqnarray}
where $\langle T \rangle$ is the mean temperature and 
$\Delta T = T - \langle T \rangle$ is the variance in
temperature. This yields the expression for \CV~\cite{landau,ben-hao, sto, shuryak, rajagopal}:
\begin{eqnarray}
\frac{1}{C} = 
\frac{(\langle T^2 \rangle -  \langle T \rangle ^2 )}  
   { \langle T \rangle ^2}.
\label{eqn4}
\end{eqnarray}
Heat capacity thus can be estimated from the fluctuations in energy or temperature. 
For a system in equilibrium, the mean values of $T$ and $E$ are
related by an equation of state. However, the fluctuations in energy
and temperature have very different behaviour.
Energy being an extensive quantity, its
fluctuation has a volume dependent component. So energy is
not suited for obtaining the heat
capacity. On the other hand, temperature fluctuations provide a good
major for estimating the \cv~\cite{landau, sto, shuryak,rajagopal}. 

The temperature of
the system can be obtained from the transverse momentum (\pT) spectra of the
emitted particles.
An exponential Boltzmann-type fit to the
\pT~spectra gives a measure of the temperature:
\begin{eqnarray}
F(p_{\rm T}) = \frac{1}{p_{\rm T}}\frac{dN}{dp_{\rm T}} \approx 
A e^{-p_{\rm  T}/T_{\rm eff}},
\label{eqn5}
\end{eqnarray}
where $A$ is a normalization factor and 
\Teff~is the apparent or effective temperature of the system~\cite{ben-hao}.
For obtaining the event-by-event fluctuation, the temperature needs to
be estimated in every event. The fitting is possible only for central
heavy-ion collisions at the LHC energies when the number of particles
is at least one thousand in every event. Even in this case, the error
associated with the fitting will be relatively large. This can be
overcome by making a connection of 
mean transverse momentum (\meanpt) of particles in every event with
the temperature. Since the calculation of the mean value is more
stable, this method of temperature estimation can also be used for
collisions at RHIC energies. The  \meanpt~can be written as~\cite{star_system}:
\begin{eqnarray}
\langle p_{\rm T} \rangle & = &
\frac
{\int_0^\infty p_{\rm T}^2  F(p_{\rm T})  dp_{\rm T}} 
{\int_0^\infty p_{\rm T}       F(p_{\rm T})  dp_{\rm T}} \\
& = & 
\frac
{2T_{\rm eff}^2 + 2m_0 T_{\rm eff} + m_0^2}
{m_0+ T_{\rm eff}},
\label{eqn6}
\end{eqnarray}
where $m_0$ is the rest mass of the particle. 
%For a massless particle:
%\begin{eqnarray}
%\langle m_{\rm T} \rangle  = \langle p_{\rm T} \rangle = 2 T_{\rm eff}
%\end{eqnarray}
Note that the
integration for \pT~is from 0 to
$\infty$. But in reality the \pT~window is finite. For a range of
\pT~within $a$ to $b$, we obtain:
\begin{eqnarray}
\langle p_{\rm T} \rangle 
&=& 
\frac
{\int_a^b p_{\rm T}^2  F(p_{\rm T})  dp_{\rm T} }
{\int_a^b p_{\rm T} F(p_{\rm T})  dp_{\rm T} } \\
&=&  
2T_{\rm eff} +  \\
& &  \frac
{a^2    e^{-a/T_{\rm eff}} -     b^2 e^{-b/T_{\rm eff}}   }
{(a+T_{\rm eff}) e^{-a/T_{\rm eff}} - (b+T_{\rm eff}) e^{-b/T_{\rm eff}}   }.
\label{eqn7}
\end{eqnarray}
This equation links the value of \meanpt~within a
specified range of \pt~to \Teff.  

\begin{figure}[tbp]
\centering \includegraphics[width=0.49\textwidth]{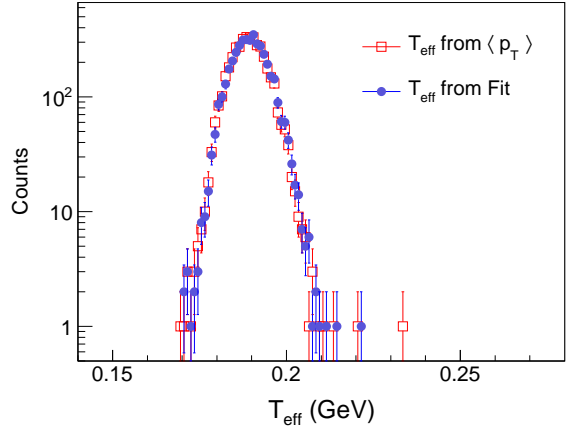} 
\caption{
Event-by-event \Teff~distributions of pions 
for central Pb+Pb collisions at 
\sNN=~2.76~TeV from the AMPT model within 
rapidity range of -1.0 to 1.0 and 
$ 0.15 < p_{\rm T} < 2.0 $~GeV. \Teff~distributions, obtained by fitting the 
\pt~distribution of each event and from the 
\meanpt~for are presented.
}
\label{fig_ampt}
\end{figure}

In order to validate the relation between \pt~to \Teff, we have
generated a large number of events using the AMPT model~\cite{ampt} 
for Pb+Pb collisions at \sNN~=~2.76~TeV. The goal is to compare the
values of \Teff~obtained from event-by-event \pt~distribution and
from \meanpt~distributions. 
For top central (top 5\% cross section) collisions,
\pt~distribution of pions has been constructed for 
each event within a rapidity range of -1.0 to 1.0. The distribution is fitted to an 
exponential function and the inverse slope parameter (\Teff) is 
extracted within fit range, $0.15 < p_{\rm T} < 2.0$~GeV. 
Fig.~\ref{fig_ampt} shows the extracted event-by-event \Teff~distribution (as solid 
circles). 
For the same set of events, the values of 
\meanpt~has been calculated within the same $\eta$ and \pt~ranges for
each event. From the value of \meanpt~for each event, the \Teff~is calculated using 
eqn.~(\ref{eqn7}).  Resulting \Teff~distribution has been plotted as open squares in 
Fig.~\ref{fig_ampt}. Both the \Teff~distributions are observed to be
same. This validates the relationship of \meanpt~and \Teff~as given in eqn.~(\ref{eqn7}). 

We note that the extracted temperature, \Teff, is a
combination of kinetic freeze-out  temperature ($T_{\rm kin}$) and transverse flow velocity
($\beta_{\rm T}$)  of the system:
\begin{eqnarray}
T_{\rm eff} = T_{\rm kin} + f(\beta_{\rm T}).
\label{eqn8}
\end{eqnarray}
For pion, $f(\beta_{\rm T}) \approx  m_0 \langle  \beta_{\rm T} \rangle^2$. 
The event-by-event fluctuations of $\beta_{\rm T}$ needs to be taken
into account for calculating the fluctuation in kinetic
temperature~\cite{teaney,bhalerao,xin-nian,Bozek}. 
Fluctuation in $\beta_{\rm T}$ dominates over the fluctuation in
\Tkin~for small systems (e.g. pp)~\cite{shuryak_beta}, 
asymmetric~(e.g., pPb)~\cite{Bozek} and non-central collisions. 
For central Au+Au collisions at \sNN~=~200~GeV~\cite{star_temp},
$\beta_{\rm T}=0.59\pm 0.051$ and for central Pb+Pb collisions at
\sNN~=~2.76~TeV~\cite{alice_temp}, $\beta_{\rm T}=0.651\pm 0.02$, which 
translate to \Tkin~as $0.095\pm 0.010$~GeV and $0.09\pm 0.005$~GeV,
respectively by using blast-wave fit~\cite{bw_fits}.
For the present work, we consider 10\% fluctuation in $\beta_{\rm T}$
and calculate its effect on specific heat. \CV~is calculated using the
equation,
\begin{eqnarray}
%\frac{1}{C_{\rm v}} & =  &
\frac{1}{C} & =  &
\frac{(\langle T_{\rm kin}^2 \rangle -  \langle T_{\rm kin} \rangle ^2 )}  
   { \langle T_{\rm kin} \rangle ^2} \approx 
\frac{(\langle T_{\rm eff}^2 \rangle -  \langle T_{\rm eff} \rangle ^2)}  
   { \langle T_{\rm kin} \rangle ^2}  \nonumber \\
& = & \frac{(\Delta T_{\rm eff})^2}   { \langle T_{\rm kin} \rangle ^2}.
\label{eqn9}
\end{eqnarray}
The values of $\langle T_{\rm kin} \rangle$ are obtained from the
blast-wave fits to the \pT~distributions of identified particles. 
With this, we obtain the specific heat as the heat capacity per
number of particles ($N$) as (\cv~=~\cvN) within the system.

Let us put the specific heat calculated in the present scenario (heat 
capacity per particle) in perspective with 
quantities normally
quoted in theoretical calculations. For an ideal gas of particles of mass $m$ and 
degeneracy factor $g$ at temperature $T$, zero chemical potential and 
volume $V$, the number of particles $N(T,V)$ can be expressed using 
Boltzmann statistics:
\bea
N & = & g \int \frac{d^3xd^3p}{h^3} \exp[-\frac{\sqrt{p^2+m^2}}{T}] \\
\nonumber
   &  = & g  \int d^3x \int \frac{d^3p}{h^3}
   \exp[-\frac{\sqrt{p^2+m^2}}{T}] \\ \nonumber
   &  = & g  \frac{VT^3}{(2\pi)^3} \int d^3q  \exp[-\sqrt{q^2+(m/T)^2}
   ] \\  \nonumber
   &  = & g  \frac{VT^3}{(2\pi)^3} \alpha, 
\eea
where $q = p/T$, $\alpha = \int d^3q  \exp[-\sqrt{q^2+(\frac{m}{T})^2} ]$ and 
we have taken $\hbar=h/(2\pi)=1$. The energy $E(T,V)$ is given by:
\bea
E = g \int \frac{d^3xd^3p}{h^3} \sqrt{p^2+m^2} \exp[-\frac{\sqrt{p^2+m^2}}{T}]. 
\eea
The heat capacity (from eqn.~\ref{eqn1}) can be written as,
\bea
C&=&g \int d^3x \int \frac{d^3p}{h^3} \biggl ( \frac{p^2+m^2}{T^2}
\biggr ) \exp[-\frac{\sqrt{p^2+m^2}}{T}] \nonumber \\
        &=& g \frac{VT^3}{(2\pi)^3} \int d^3q (q^2 + (\frac{m}{T})^2)
         \exp[-\sqrt{q^2+(\frac{m}{T})^2} ] \nonumber \\
        &=& g \frac{VT^3}{(2\pi)^3} \beta.
\eea
$\beta = \int d^3q (q^2 + (\frac{m}{T})^2) \exp[-\sqrt{q^2+(\frac{m}{T})^2} ] $
is a dimensionless quantity. The specific heat is the heat capacity
per unit phase space volume, 
\bea
c_{\rm v}=C/\Delta,
\eea
where $\Delta$ is an  estimate of the phase space volume.
In lattice calculations one extracts the dimensionless 
quantity \cvT~and investigate its temperature
dependence~\cite{swagato}, so in these calculations $\Delta = VT^3$.
However, in 
experiments it is simpler to measure the dimensionless quantity $C/N$ where $N$ is the 
charged particle multiplicity, and thus $\Delta = N$, where $N$ is
taken as pseudorapidity~($\eta$) density of charged particles at mid rapidity
($dN_{\rm ch}/d\eta$ at $\eta=0$).
We compare the experimental results to 
other model calculations for $C/N$ as in Ref.~\cite{ben-hao},
where a parton and hadron cascade model, PACIAE has been used to compute $C/N$. We also 
compare with HRG where it is straightforward to obtain both, \cvT~and $C/N$. 

\section{Hadron Resonance Gas Model and Specific Heat}\label{sec.hrg}

The current continuum estimates of lattice QCD thermodynamics in the 
low temperature and density phase (hadronic phase) show good agreement 
with that of an ideal hadron resonance gas~\cite{LQCDHRG1, LQCDHRG2, LQCDHRG3}. 
Assuming complete chemical equilibrium between all hadrons, 
the hadron chemical potential of the $i^{th}$ species $\mu_i$ can be written as 
\be 
\mu_i = B_i\mu_B + Q_i\mu_Q + S_i\mu_S,
\label{eq.mui}
\ee 
where $B_i$, $Q_i$ and $S_i$ are the 
baryon number, electric charge 
and strangeness quantum numbers of the $i^{th}$ hadron, 
respectively. 
The HRG partition function $Z$ in 
the grand canonical ensemble at $\left( T,\mu_B,\mu_S,\mu_Q  \right) $
can be expressed as:
\begin{eqnarray}
\ln &Z&= \nonumber \\    
&& VT^3\sum_i\frac{g_i}{2\pi^2}\left(\frac{m_i}{T}\right)^2 
 \sum_{l=1}^{\infty}\left(-a\right)^{l+1}l^{-2}K_2\left(lm_i/T\right) 
   \nonumber \\
&& \exp[l\left(B_i\mu_B+Q_i\mu_Q+S_i\mu_S\right)/T],
\label{eq.pressureHRG}
\end{eqnarray}
where the sum runs over all hadrons and resonances up to mass $\sim$2~GeV as listed 
in the Review of Particle Physics~\cite{PDG}. $a=-1$ for mesons and $1$ for baryons. 
Here $g_i$, $m_i$, $B_i$, $Q_i$ and $S_i$ refer to the properties of
the $i$th hadron species:
its degeneracy factor, mass, baryon 
number, electric charge and strangeness respectively. $V$ is the volume of the 
fireball under study. $K_2$ is the modified Bessel function of the second kind. 
From $\ln Z$, all thermodynamic quantities could be computed. 

From $\ln Z$, $E$ is obtained as follows:
\bea 
E &=& T^2\frac{\partial \ln Z}{\partial T} + \sum_i \mu_iN_i\label{eq.energy}
\eea 
where,
\bea 
N_i &=& T\frac{\partial \ln Z}{\partial\mu_i}\label{eq.number}. 
\eea 
From $E$ it is straightforward to compute $C$ using Eq.~\ref{eqn1}.

\begin{figure}[tbp]
\centering \includegraphics[width=0.49\textwidth]{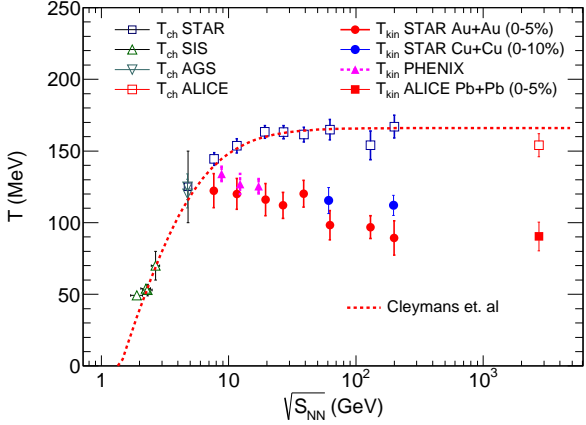} 
\caption{Chemical and kinetic freeze-out temperatures for
  central Au+Au~\cite{star_temp} and Cu+Cu~\cite{star_system}
  collisions at RHIC energies, 
  and Pb+Pb collisions at \sNN~=~2.76~TeV~\cite{alice_temp}. Thermal
  model calculation~\cite{cleymans} to $T_{\rm ch}$ is also shown. 
}
\label{fig_tch_tkin}
\end{figure}

The values of \Tkin~have been reported by experiments at RHIC and LHC,
as the final state particles give the information about \Tkin~from the 
particle spectra~\cite{sandeep_freeze}. 
These are obtained by making combined fits to the 
identified particle spectra using the Boltzmann-Gibbs blast-wave model.
Figure~\ref{fig_tch_tkin} gives the 
\Tkin~values for different beam 
energies and collision systems~\cite{star_temp,star_system,alice_temp}. 
In addition, chemical freeze-out temperatures (\Tch), extracted from the 
identified particle yield by using thermal model 
calculations~\cite{cleymans, anton}, are also shown in the figure. 
We find that the difference between \Tch~and \Tkin~increases with the increase 
of beam energy. Lattice calculations indicate that \cv~is a monotonically 
increasing function of $T$ at zero $\mu_B$. Thus we expect that the difference between the 
\cv~ extracted at the chemical and kinetic freeze-out surfaces should also increase 
with beam energy following the trends of \Tch~and \Tkin.

\begin{figure}[tbp]
\centering \includegraphics[width=0.49\textwidth]{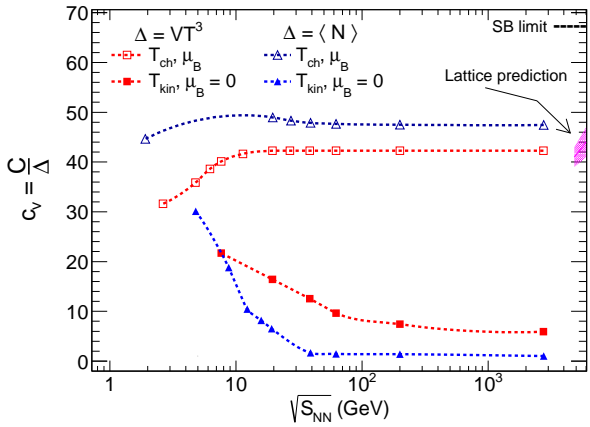} 
\caption{
  Specific heat, \cv~as a function of collision 
  energy from the HRG model for two temperature settings, \Tch~with finite 
  $\mu_{\rm B}$,  and \Tkin, and two
  phase space volumes, $\Delta=VT^3$ and  $\langle N \rangle$. Lattice
  calculation for \cv~at $T=154\pm 9$~MeV~\cite{swagato} and Stefan-Boltzmann limit are
  indicated in the figure. 
}
\label{Specific1}
\end{figure}

We calculate \cv~from HRG model for two scenarios. In the first case, we compute at 
the chemical freeze-out surface using the extracted \Tch~as well as $\mu_B$. However, in the 
experiment one can only determine the \cv~at the kinetic freeze-out surface where the 
momentum exchange freezes. The thermal conditions at the kinetic freeze-out surface are much 
different from that at the chemical freeze-out surface. Hence the fireball is expected to have 
different \cv~at the two surfaces. In the second case, we try to estimate \cv~at the kinetic 
freeze-out surface using \Tkin and zero hadron chemical potentials. For both scenarios, we calculate 
\cvT~and \cvN~for a wide range of beam energy, from \sNN~=~1.91~GeV to 2.76~TeV.
The results of \cv~are shown in the Fig.~\ref{Specific1}. 
It is observed that the trend of 
\cv~as a function of \sNN~is similar to 
the nature followed by chemical and kinetic freeze-out
temperatures. The value of
\cvN~corresponding to \Tkin~shows a sharp drop with increase of
energy, and beyond \sNN~=~62.4~GeV, the rate of decrease is very slow. A better estimate 
of \cv~in HRG could be made with realistic hadron chemical potentials taking into account 
the conservation of hadron number from the chemical to the kinetic freeze-out surfaces.

Recent lattice calculations for \cv~have been reported 
\cite{swagato} as a function of temperature. The lattice results are 
at zero baryonic potential, hence only relevant at the LHC and higher energies. The 
value of \cv~as indicated in the Fig.~\ref{Specific1} is for
$T=154\pm 9$~MeV, corresponding to the QCD transition temperature. 
It is seen that at the transition temperature and below, 
HRG results of \cvT~agree well with lattice calculations~\cite{swagato}. 
The Steffan-Boltzmann non-interacting gas limit (\cv~$\approx$~66) is
also shown in the figure.

\section{Distributions of \meanpt~and \Teff}

\begin{figure*}[tbp]
\centering \includegraphics[width=0.311\textwidth]{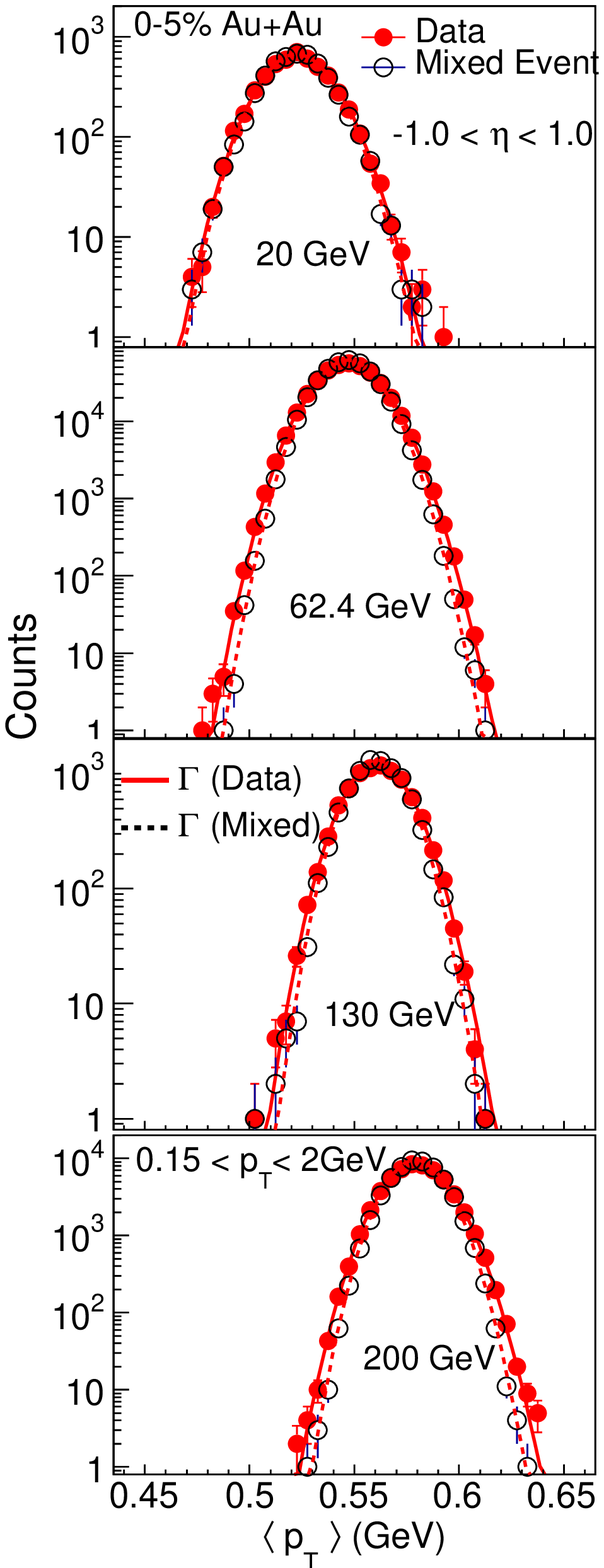} 
\centering \includegraphics[width=0.31\textwidth]{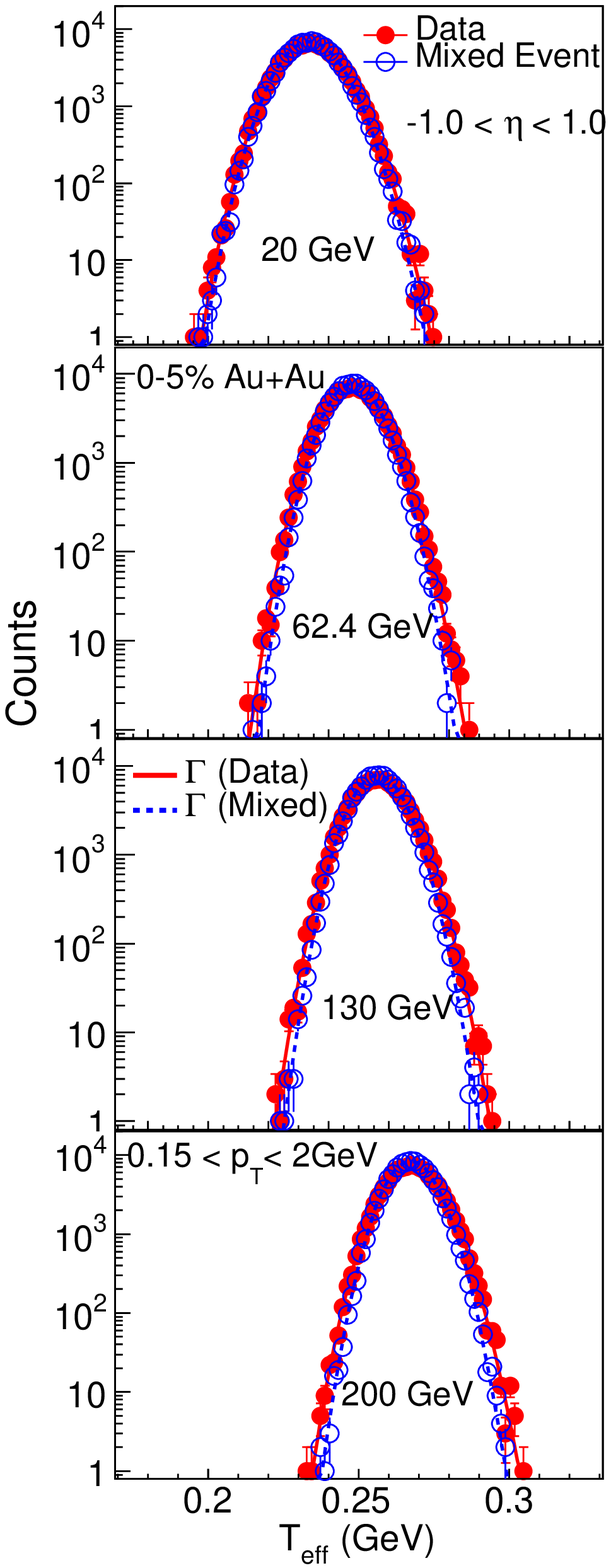} 
\caption{
Left panels show event-by-event mean transverse momentum 
distributions for 5\% most central 
Au+Au collisions at \sNN~=~20, 62.4, 130 and 200~GeV within 
$|\eta | < 1 $ and $ 0.15 < p_{\rm T} < 2.0 $~GeV~\cite{star2}. 
Distributions for mixed events are superimposed on 
the data. The solid and dashed lines show the fits with $\Gamma$
functions. The right panels show the extracted \Teff~distributions 
for each incident energy.  }
\label{fig_AuAu}
\end{figure*}

Experimental data for \meanpt~distributions have been reported by
experiments at SPS and
RHIC~\cite{NA49-1,NA49-2,NA49-3,CERES,phenix,star1,star2}. 
In the left panel of Fig.~\ref{fig_AuAu} we present the
\meanpt~distributions from the STAR experiment~\cite{star1,star2} 
for the 5\% most central Au+Au collisions at \sNN~=~20, 62.4, 130 and 200~GeV. The
results are shown for charged particle tracks within 
$|\eta | < 1 $ and $ 0.15 < p_{\rm T} < 2.0 $~GeV.
The solid points are the event-by-event \meanpt~distributions from the experimental
data, whereas the open circles are the corresponding results for mixed
events. The mixed events are created by randomly selecting charged
particles from different events. The mixed event distributions contain
all the systematic effects arising from the detector effects,
such as efficiency and acceptance, as well as include statistical
fluctuations. The non-statistical or dynamical fluctuations in
\meanpt~can be extracted by subtracting the width
of the mixed event distribution from that of the real data. 

\begin{figure*}[tbp]
\centering \includegraphics[width=0.31\textwidth]{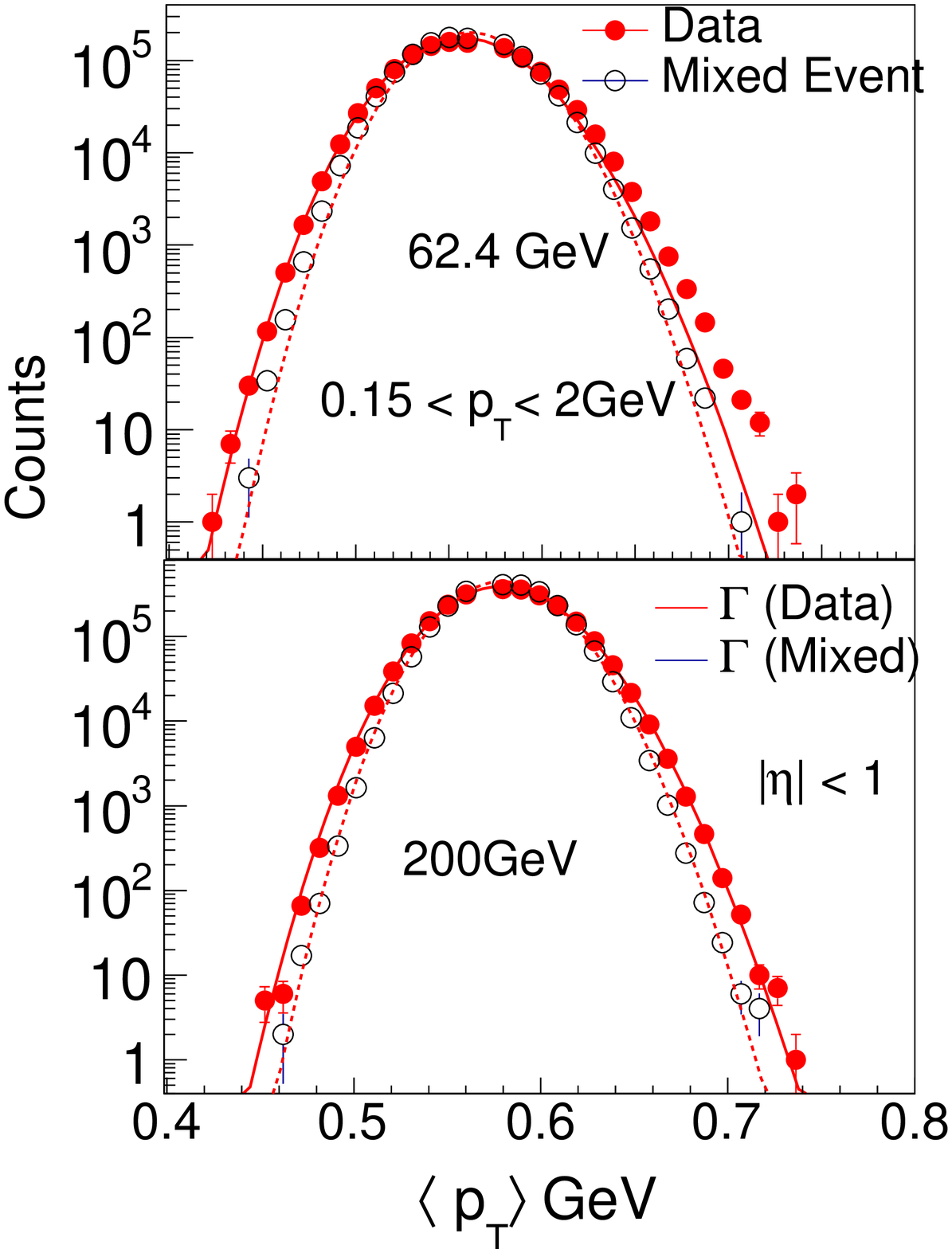} 
\centering \includegraphics[width=0.31\textwidth]{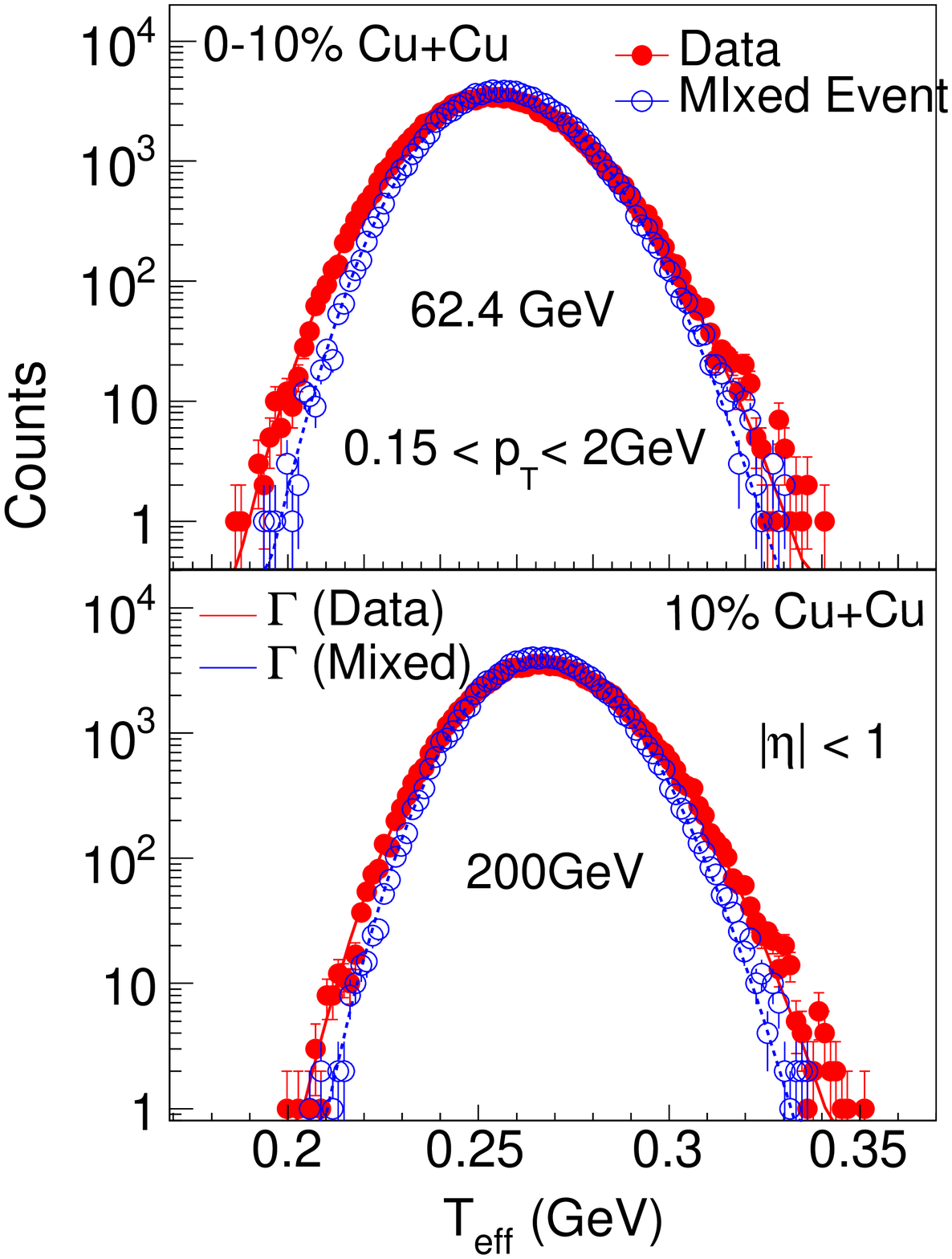} 
\caption{
Similar distributions as in Fig.~\ref{fig_AuAu} for 
10\% most central 
Cu+Cu collisions at \sNN~=~62.4 and 200~GeV~\cite{star_system}. 
 }
\label{fig_CuCu}
\end{figure*}

It has been observed that the \meanpt~distributions are nicely
described by using the gamma ($\Gamma$)
distribution~\cite{star1,star2,mike2}:
\begin{eqnarray}
f(x) = \frac{x^{\alpha-1}e^{-x/\beta}}{\Gamma(\alpha)\beta^\alpha}.
\end{eqnarray}
Here $x$ represents the \meanpt. The mean ($\mu$) and standard
deviation ($\sigma$) of the distribution are related to the fit parameters ($\alpha$ and
$\beta$) by $\mu = \alpha\beta$ and $\sigma = \sqrt{\alpha\beta^2}$.
Both the real and mixed event \meanpt~distributions are fitted with
the $\Gamma$ function and the fits are shown by
the solid and dashed lines, respectively, in
the left panels of Fig.~\ref{fig_AuAu}.
The fitted distributions are used to generate a large
number of \meanpt~values  for which corresponding \Teff~values
are calculated from eqn.~\ref{eqn7}. 
The resulting histograms represent event-by-event \Teff~distributions,
which are shown in the right panels of
Fig.~\ref{fig_AuAu} for both real data and mixed events.
These distributions are also fitted by the $\Gamma$ function as shown
by the solid 
and dashed lines for data and mixed events, respectively. Table~\ref{table1}
lists the fit parameters for event-by-event
\Teff~distributions for data and mixed events.

The system size dependence of \meanpt~and \Teff~distributions 
have been studied with the STAR experimental data of top 10\% central 
Cu+Cu collisions 
at \sNN~=~62.4 and 200 GeV~\cite{star_system}. The results are 
presented in in Fig.~\ref{fig_CuCu}. Corresponding 
$\Gamma$ distribution fit parameters to the event-by-event 
\Teff~distributions for top 10\% central collisions are tabulated in Table~\ref{table2}.

\begin{table}[htbp]
\caption{
The event-by-event \Teff~distributions for central (top 5\%) Au+Au 
collisions are fitted by the gamma function. Table gives fhe fit parameters,
$\alpha$ and $\beta$ along with mean ($\mu$) and standard deviation ($\sigma$).
}
\begin{tabular}{|c|c|c|c|c|c|} 
\hline 
$\sqrtsign s_{\rm NN}$ & Case & $\alpha$    & $\beta$                 & $\mu$         &  $\sigma$ \\  
(GeV)     &    &      & (GeV) &  (GeV)  & (GeV)\\ \hline 
20     & data    & 658.53      &   3.556$\times$10$^{-4}$   &   0.2341    &  0.00912 \\
20     & mixed & 724.56      &   3.229$\times$10$^{-4}$     &   0.2339    &  0.00869 \\\hline 
62.4     & data    & 860.20      &  2.885$\times$10$^{-4}$     &   0.2482    &  0.00846 \\ 
62.4    & mixed &  1043.67    &   2.378$\times$10$^{-4}$     &   0.2481    &  0.00768 \\\hline 
130   & data    & 920.25      &   2.789$\times$10$^{-4}$     &   0.2566    &  0.00846 \\
130     & mixed & 1140.12  &   2.249$\times$10$^{-4}$     &   0.2564    &  0.00759 \\ \hline 
200     & data    & 1078.23    &  2.483$\times$10$^{-4}$     &   0.2677    &  0.00815 \\
200     & mixed & 1387.56    &  1.927$\times$10$^{-4}$    &   0.2674   &  0.00718 \\
\hline 
\end{tabular}
\label{table1}
\end{table}
\begin{table}[htpb]
\caption{
The event-by-event \Teff~distributions for central (top 10\%) Cu+Cu 
collisions are fitted by the gamma function. Table gives th fit parameters,
$\alpha$ and $\beta$ along with mean ($\mu$) and standard deviation ($\sigma$).
}
\begin{tabular}{|c|c|c|c|c|c|} 
\hline 
$\sqrtsign s_{\rm NN}$ & Case & $\alpha$    & $\beta$                 & $\mu$         &  $\sigma$ \\  
(GeV)     &    &      & (GeV) &  (GeV)  & (GeV)\\ \hline 
62.4     & data    & 211.88      &  12.040$\times$10$^{-4}$     &   0.2550    &  0.0175 \\ 
62.4    & mixed &  271.94    &    9.455$\times$10$^{-4}$     &   0.2571   &  0.0156 \\\hline 
200     & data    & 277.08  &    9.687$\times$10$^{-4}$     &   0.2684    &  0.0161 \\
200     & mixed & 370.71   &  7.278$\times$10$^{-4}$    &   0.2698   &  0.0140 \\
\hline 
\end{tabular}
\label{table2}
\end{table}

From these two figures and the given tables for \meanpt~and
\Teff~distributions for Au+Au and Cu+Cu collisions at RHIC energies, we can
infer that: (a) the mean values of the event-by-event \meanpt~and \Teff~consistently
increase with the increase of beam energy, (b) the widths of the
distributions decrease with the increase of beam energy. 
In addition, the widths for Cu+Cu system are observed to be larger
compared to the corresponding widths of the Au+Au system. This may be
because of the smaller system size for Cu+Cu compared to Au+Au
system. 

Experimental data for event-by-event \meanpt~distributions are not
available for Pb+Pb collisions at LHC energies~\cite{alice_meanpt}. 
The string melting mode of
AMPT model is used to
generate central (top 5\%) Pb+Pb collisions at \sNN~=~2.76~TeV.
The \meanpt~and \Teff~distributions
are constructed from these generated events as shown in Fig~1.
This distribution will be used to extract specific heat at the LHC energy.

\section{Specific heat from experimental data}

The widths of the \Teff~distributions are 
strongly affected by statistical fluctuations, which need to be 
subtracted as the heat capacity is related only to the dynamical part 
of the fluctuation. The width contains two components:
\begin{equation}
 (\Delta T_{\rm eff})^2 = (\Delta T^{dyn}_{\rm eff})^2 + (\Delta T^{stat}_{\rm eff})^2. 
\end{equation}
$\Delta T^{dyn}_{\rm eff}$ values are obtained by subtracting the widths of the \Teff~distributions for 
mixed events from the real data. With this, eqn.~\ref{eqn9} is expressed as:
\begin{equation}
\frac{1}{C}  =  \frac{(\Delta T^{dyn}_{\rm eff})^2}   {\langle
  T_{\rm kin} \rangle ^2}. 
\label{eqnfinal}
\end{equation}

The heat capacity \CV~is calculated from eqn.~\ref{eqnfinal} by using 
the values of \Tkin~from Fig.~\ref{fig_tch_tkin}. Knowing the heat 
capacity, the specific heat, \cv~is obtained by dividing \CV~by 
number of 
charged particles in the system. Since the experimental 
results presented here are at mid-rapidity, we have 
divided the value of \CV~by charged particle multiplicity 
at mid-rapidity~\cite{nch1,nch2} to obtain the specific 
heat. This is presented in Fig.~\ref{Specific1} for Au+Au and Cu+Cu collisions 
at RHIC energies. The estimated $C/N$ for the LHC energy from 
the AMPT model using Fig.~\ref{fig_ampt} is also 
shown in the figure. 
The errors in the data points are estimated mainly from the following sources:
(a) error in extraction of \Tkin~using the blast-wave fits,
(b) error in charge particle multiplicity density,
and (c) error in \meanpt~as reported in the experimental data. The 
error in \Tkin~takes into account the spread in the value of ($\beta_{\rm T}$). 
It is observed that $C/N$ has a sharp drop from \sNN~=~20~GeV to 62.4~GeV, beyond
which the decrease is rather slow up to the LHC energy.

\begin{figure}[tbp]
\centering \includegraphics[width=0.49\textwidth]{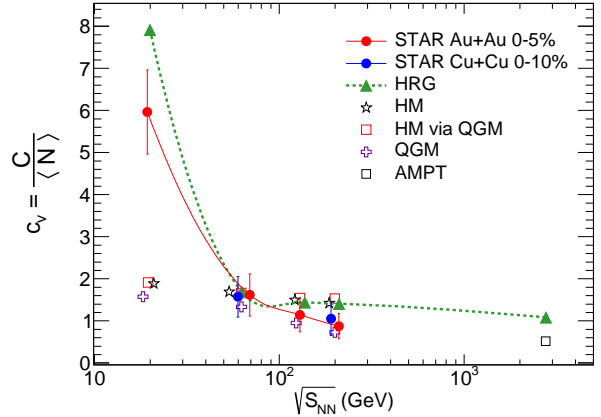} 
\caption{
  Specific heat, \cv~as a function of collision
  energy for central Au+Au and Cu+Cu collisions at RHIC
  energies. Result from AMPT model is given for the LHC energy at \sNN~=~2.76~TeV.
  HRG calculations at \Tkin~are shown in the figure. 
  Model calculations for three different scenarios from Ref.~\cite{ben-hao} are superimposed 
  on the experimental results. 
}
\label{Specific2}
\end{figure}

HRG calculations for \cvN~with \Tkin~are superimposed in
Fig.~\ref{Specific2}. These results follow the experimental data
points quite well. 
In Ref.~\cite{ben-hao}, specific heat for central (top 5\%) Au+Au 
collisions at RHIC energies are discussed using 
a parton and hadron cascade model, called 
PACIAE. The results of the model calculations are presented for three 
cases: hadronic matter in the final state (HM), quark-gluon matter in the 
partonic initial state (QGM), and hadronic matter via quark-gluon 
matter (HM via QGM). These results for Au+Au collisions are also 
presented in Fig.~\ref{Specific2}. 
The results of these models miss the experimental
data point at \sNN~=~20~GeV, but can explain the data 
at higher energies. The PACIAE is a transport model with LO pQCD 
cross sections considered between partons. It might be possible 
that at lower beam energies, the LO pQCD description is not appropriate.

\section{Discussion}

In an earlier publication by R. Korus {\it et al.} (Ref.~\cite{korus}), the experimental data
of \pT~correlations from the NA49~experiment~\cite{NA49-1,NA49-2,NA49-3} for Pb+Pb collision at
laboratory energy of 158~GeV had  used to calculate specific heat, which yielded a value
of $60\pm 100$. The large
error bars of these results made the reported results insignificant.
One of the possible reasons for the large errors is the low
particle multiplicity which gives a significant hindrance to the
calculation of dynamic temperature fluctuations~\cite{voloshin}.
At the SPS energies, \meanpt~distributions have been reported by 
NA49 collaboration for laboratory energies of 20, 30, 40 80 and
158~GeV~\cite{NA49-1,NA49-2,NA49-3} and the CERES collaboration at 40, 80 and
158~GeV~\cite{CERES}. 
In both of the data sets, the \meanpt~distributions for
real data and mixed event are indistinguishable. These are also
prominent in the ratio plots of real and mixed events as shown in
Ref.~\cite{CERES}. Thus the extraction of dynamic fluctuation in
temperature and so the specific heat is not possible. In the present work,
we have probed much higher energy collisions, where the charged
particle multiplicities in each event are large, allowing for the 
extraction of the dynamical part of the temperature fluctuation by
overcoming the statistical fluctuations. 

% Theoretical calculations based on lattice and different models depend
% on at which temperature specific heat is evaluated. 
% Most of these calculations are done at \Tch~, whereas in experiment
% the \cvT~are calculated from at \Tkin~. 
% Figure~\ref{fig_tch_tkin} and \cvT~results from HRG model as shown in 
% Fig.~\ref{Specific2} show similar trends for \Tch and \Tkin~as well
% as \cvT~obtained from \Tch~and \Tkin.

%At zero $\mu_B$, lattice computations suggest that the 
%specific heat is a monotonically increasing function of $T$. On comparing 
%Figs.~\ref{fig_tch_tkin} and \ref{Specific1}, we find similar trends in 
%the extracted values of the specific heat from data.

The results of specific heat for Cu+Cu collisions 
are close to that of Au+Au
collisions. This shows that although a large change of volume happens
in going from Cu+Cu to Au+Au systems, the two systems are not very
different thermodynamically.

Several sources of uncertainty may affect the extraction of specific
heat. In the context of the results presented in Fig.~\ref{Specific2}, three sources
of uncertainty, {\it viz.}, effect of finite particle multiplicity, spread
of \Tkin from the fits of \pT distributions, and the radial flow
fluctuations, have been discussed. Apart from these, it is worthwhile
to point out some other uncertainties.  Fluctuations in the impact
parameter of the collision and thus the fluctuation in the number of
participating nucleons gives an uncertainty to the event-by-event mean
\pT distribution. Choice of narrow bins in centrality has been made to
minimize this uncertainty. Another source of uncertainty may come from
the choice of the \pT window.  The lower bound of the \pT window needs
to be chosen properly in order to reduce the final state effects such
as resonance decay and hadronic scattering. Similarly, the upper limit
on\ pT needs to be chosen such that the effect from mini-jets and jets
are minimized. Although it is implicit that radial flow fluctuations
are minimum for central collisions because of inherent symmetry of the
system, its detailed study can be made using an event-by-event
hydrodynamic model. These studies will help to pin down the errors of
the results extracted in Fig.~\ref{Specific2}.

\section{Summary}

 We have studied the excitation energy dependence of specific
 heat of hadronic matter formed in heavy-ion collisions corresponding 
 to RHIC and LHC energies. 
 In the present work, dynamical component of the temperature 
 fluctuation is calculated from \meanpt~distributions. From this, the specific
 heat is obtained as heat capacity per charge particle.
 We employ the HRG model to calculate heat  
 capacity from the variation of energy of the system with 
 temperature. Results of the HRG calculations are close to
 the data. With increase of collision energy,
 \cv~shows a sharp drop from low energy till
 \sNN~=~62.4~GeV, beyond which the rate of decrease is very slow.
  In this regard, we look forward to results of BES program of RHIC, where the collision energy and 
  centrality dependences of \cv~are expected to provide
  important signatures for the onset of the QGP phase
  transition.  In order to probe the QCD critical point, we
   propose a finer scan of beam energies for the second
  phase of BES program (BES-II) from 7.7~GeV to 62.4~GeV. A sudden change in \cv~is
  expected at a particular beam energy within this range.
  Studies of heat capacity at high baryon density and lower
  temperatures accessible at Facility for Antiproton and Ion
  Research (FAIR) would be of high interest, unless it is critically
  challenged by statistical fluctuations. 
  Predictions for \cv~at the LHC at
  \sNN~=~2.76~TeV are made using different models.
  It will be interesting to obtain \cv~at the highest LHC 
  energy of \sNN~=~5.02~TeV in order to make a direct comparison to 
  lattice calculations. In literature, 
  it has been also proposed to calculate thermal conductivity 
  from transverse energy ($E_{\rm T}$) fluctuations, which can be explored in future
  studies. The excitation energy dependence of \cv~provides important 
  information regarding the  thermodynamic properties, such as,
  heat conductivity, speed of sound ($c_{\rm s}^2$), compressibility ($k_{\rm T}$),
  etc., which may reveal better understandings of the matter formed in 
  relativistic nuclear collisions.

\noindent
{\bf Acknowledgement}
SC acknowledges ``Centre for Nuclear Theory" [PIC XII-R$\&$D-VEC-5.02.0500], Variable Energy 
Cyclotron Centre for support. This research used resources of the  LHC grid computing centre 
at the Variable Energy Cyclotron Centre, Kolkata.

\end{document}